\documentclass[global,twocolumn]{svjour}
\usepackage{graphicx}
\usepackage{siunitx}
\usepackage{gensymb}
\usepackage{xcolor} 
\journalname{myjournal}
\begin{document}
	\sloppy
%
\title{Dosimetric characterisation and application to radiation biology of a kHz laser-driven electron beam}
\author{Marco Cavallone\inst{1} \and Lucas Rovige\inst{1} \and Julius Huijts\inst{1} \and \'Emilie Bayart\inst{1,2} \and Rachel Delorme\inst{4} \and Aline Vernier\inst{1}\and Patrik Gon\c calves Jorge\inst{5} \and Rapha\"{e}l Moeckli\inst{5} \and Eric Deutsch\inst{3} \and Jér\^{o}me Faure\inst{1} \and Alessandro Flacco\inst{1}
\thanks{\emph{Corresponding author:} alessandro.flacco@ensta-paris.fr}%
}                     
\offprints{}          
\institute{Laboratoire d’Optique Appliqu\'ee, ENSTA Paris, \'Ecole Polytechnique, CNRS-UMR7639, Institut Polytechnique de Paris,  91762 Palaiseau cedex, France \and SFRO - RadioTransNet, Centre Antoine Béclère, 47 rue de la Colonie, 75013 Paris \and INSERM 1030, Univ Paris-Sud, Univ Paris-Saclay, Department of Radiation Oncology, Gustave Roussy Cancer Campus, Villejuif, France \and Univ. Grenoble Alpes, CNRS, Grenoble INP, LPSC-IN2P3, 38000 Grenoble, France \and Institute of Radiation Physics, Lausanne University Hospital, Lausanne, Switzerland}
\date{Received: date / Revised version: date}
%
\maketitle
\begin{abstract}
Laser-plasma accelerators can produce ultra short electron bunches in the femtosecond to picosecond duration range, resulting in high peak dose rates in comparison with clinical accelerators. This peculiar characteristic motivates their application to radiation biology studies to elucidate the effect of the high peak dose rate on the biological response of living cells, which is still being debated. Electron beams driven by kHz laser systems may represent an attractive option for such applications, since the high repetition rate can boost the mean dose rate and improve the stability of the delivered dose in comparison with J-class laser accelerators running at \SI{10}{Hz}. In this work, we present the dosimetric characterisation of a kHz, low energy laser-driven electron source and preliminary results on \textit{in-vitro} irradiation of cancer cells. A shot-to-shot dosimetry protocol enabled to monitor the beam stability and the irradiation conditions for each cell sample. Results of survival assays on HCT116 colorectal cancer cells are in good agreement with previous findings reported in literature and validate the robustness of the dosimetry and irradiation protocol.

\end{abstract}
\section{Introduction}
\label{intro}
\sloppy
Laser-plasma accelerators can produce ultra short electron bunches with duration in the range from femtosecond to picosecond. Both quasi-monoenergetic high-energy electrons (hundreds of MeV)~\cite{mirzaie_effect_2018,faure_experiments_2008,thaury_shock_2015} and high-charge, low-energy (few MeV) electrons with broadband spectrum~\cite{guillaume_physics_2015} have been obtained with 100 TW class lasers. The peculiarity of these sources is the ultra-high peak dose rate in the pulse (up to \SI{e10}{Gy/s}) that is orders of magnitude higher than conventional Linac (\({\sim\SI{e2}{Gy/s}}\))~\cite{ashraf_dosimetry_2020}. The impact of electrons delivered in such short pulses of ultra-high dose rates on living cells or tissues still need to be extensively explored. Recent results obtained with FLASH-RT~\cite{favaudon_ultrahigh_2014,vozenin_advantage_2019,vozenin_biological_2019} and laser-accelerated protons~\cite{raschke_ultra-short_2016,bayart_fast_2019} indicate that the temporal structure of high dose rate, pulsed irradiation may have an impact on the radiobiological response and ask for a deeper understanding of the phenomena triggered by high dose rate irradiation.\\ To date, Laser-Accelerated Electrons (LAEs) produced by commercially available J-class lasers running at 10 Hz have been used for radiation biology studies~\cite{andreassi_radiobiological_2016,laschinsky_radiobiological_2012,oppelt_comparison_2015}. The main limit of these beams is the shot-to-shot pointing fluctuation in the order of the beam divergence, that hinders the reach of the stability standards required in clinics. In this context, low-energy electrons (few MeV) driven by kHz lasers may represent an interesting alternative for radiobiology applications. Such laser-plasma accelerators have been developed in recent years~\cite{he_high_2013,salehi_mev_2017,guenot_relativistic_2017,faure_review_2019} and, to our knowledge, no radiobiology study in these conditions is reported in literature at the time of writing. The key asset of these beams is the high repetition rate, which boosts the mean beam current and allows averaging of the shot-to-shot fluctuations by integration of a large number of shots. From a dosimetric point of view, this translates into higher mean dose rates (Gy/s), compared to those obtained with {J-class} lasers (Gy/min), and into a higher stability of the dose distribution at the sample. \\In this article we present the dosimetric characterisation of a low-energy kHz laser-driven electron beam for radiation biology applications and report preliminary results on \textit{in-vitro} irradiation of HCT116 colorectal cancer cells. The dose distribution at the irradiated sample was measured at each irradiation with absolutely calibrated radiochromic films, which have been shown to be dose-rate independent over a wide range of dose rate up to \SI{e12}{Gy/s}~\cite{karsch_dose_2012,jaccard_high_2017,bazalova-carter_comparison_2015}. The use of an IBA Razor Nano Chamber provided a real time dose monitoring during irradiations and evaluation of the dose uncertainty in the post-processing analysis. 

\section{Electron source and set-up}
\label{source and set up}
\begin{figure}[]
	\centering
	\includegraphics[width=0.9\columnwidth]{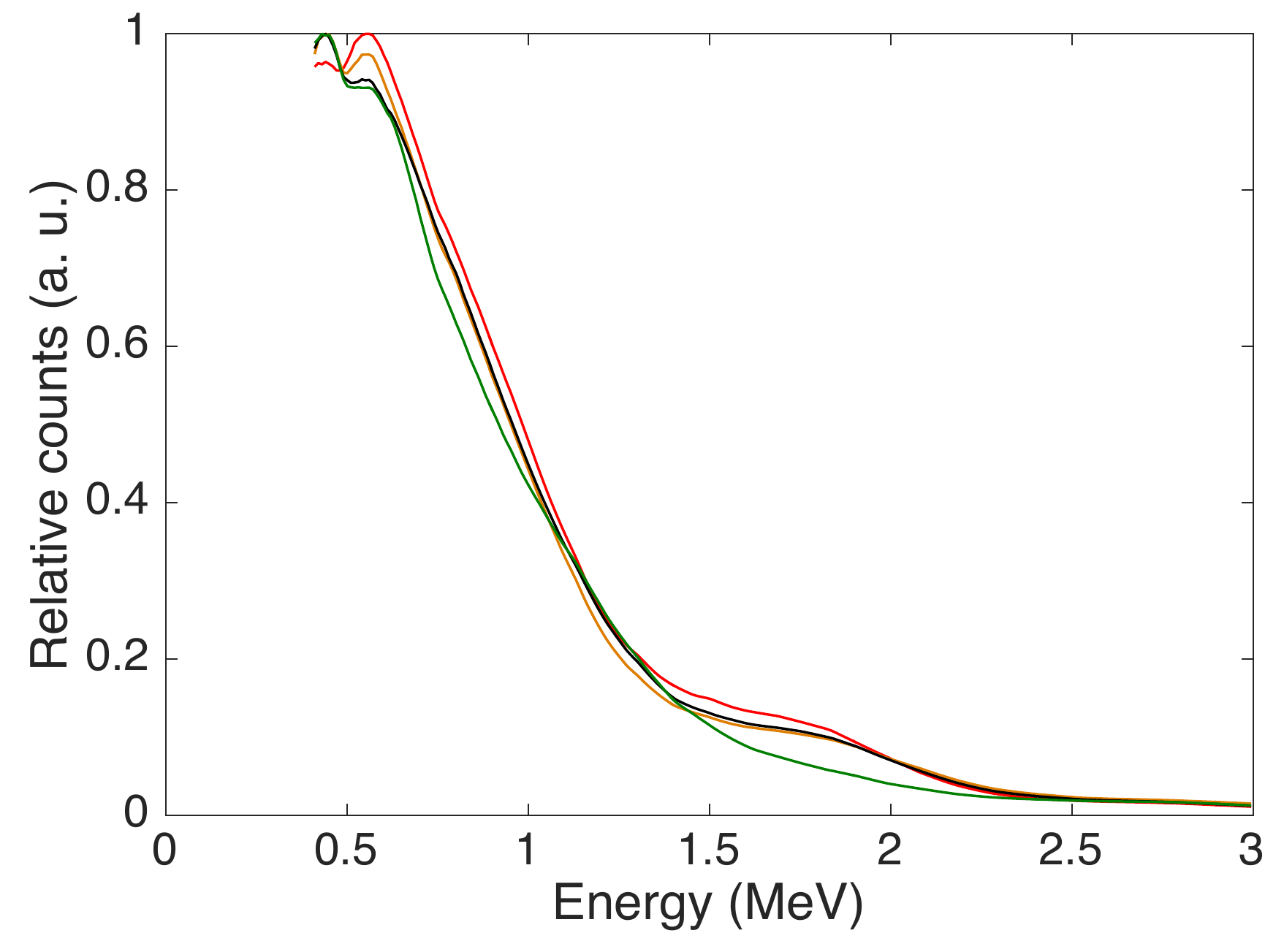}
	\caption{Spectrum of the electron beam taken in four different days, obtained by integrating 15 shots. The lower limit of the spectrometer is \SI{400}{keV}.}
	\label{spectra}
\end{figure}

\sloppy
The experiment was performed with the \textit{Salle Noire} laser system of the \textit{Laboratoire d'Optique Appliquée}, delivering 3.5 fs pulses at 1 kHz repetition rate with \SI{3}{mJ} on target. The laser can generate low energy electrons in a reproducible way in a range from sub-MeV up to few MeV with charges of \(\sim\)\SI{10}{pC} and few pC respectively, depending on the  gas density and profile~\cite{gustas_high-charge_2018}. In the experiment, the laser pulse was focused with a parabola (\SI{50}{mm} focal length) onto an N\(_2\) gas-jet. The gas-jet density was optimised so as to maximise the accelerated charge per bunch, which corresponds to the production of an electron beam featuring a low-energy quasi-thermal spectrum up to \(\sim\)\SI{2}{MeV} and a divergence of \SI{70}{mrad}.
\begin{figure}[]
	\includegraphics[width=1\columnwidth]{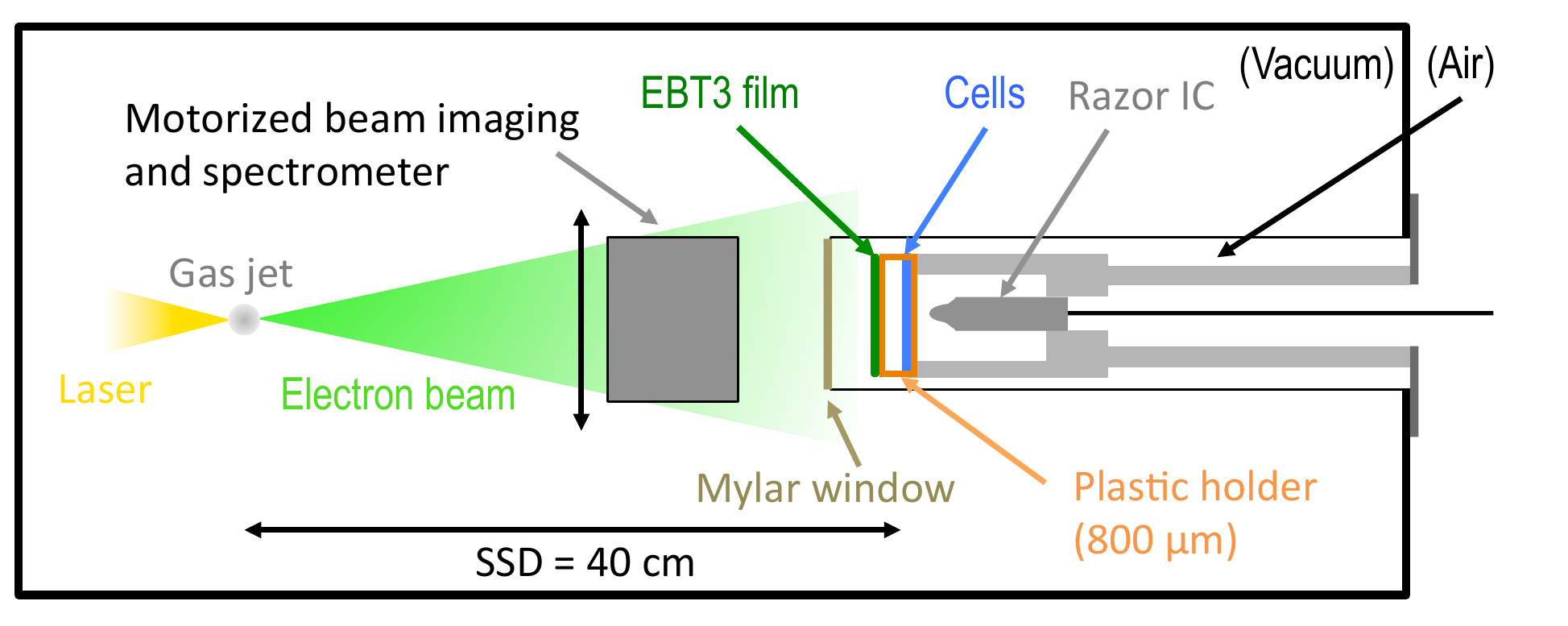}
	\caption{Schematic draft (horizontal cut, top view) of the experimental set-up.}
	\label{set-up_sallenoire}
\end{figure}
Fig.~\ref{spectra} shows the electron spectrum taken in four different days. As shown, the spectral shape can be reproduced from one day to another with good precision. The measured charge (total charge at the source) was \(9.8\)~pC/shot with an instability of \(\pm0.7\)~pC/shot (standard deviation).\\A draft of the experimental setup is shown in Fig.~\ref{set-up_sallenoire}. A tube was inserted in the vacuum chamber to place the irradiation site (in air) closer to the electron source, compatibly with the beam diagnostics. A \SI{100}{\micro\meter} thick vacuum-air Mylar window was placed \SI{38}{\centi\meter} far from the source. At this location, the transverse dimension of the beam was \SI{3}{\centi\meter} (FWHM), which allowed to irradiate a circular spot of  \SI{1}{\centi\meter} diameter with a sufficiently homogeneous dose. The cell sample was positioned vertically in air, \SI{2}{\centi\meter} after the Mylar window, with a \SI{800}{\micro\meter} thick plastic holder aligned to the beam-axis. 
A customised metallic support allowed to hold in place the cell holder and to place it at the same depth in the tube at each irradiation. The support was also designed to house an IBA Razor Nano Chamber (RNC), which was inserted behind the cell sample. The chamber (\SI{2}{\milli\meter} outer electrode diameter, \SI{0.003}{\centi\meter\squared} active volume) was used as on-line dose monitor during irradiations and as day-to-day dose reference during the optimisation of the interaction conditions between the laser and the gas-jet prior to an experimental run. A radiochromic EBT3 film was placed on the front side (in beam's view) of the cell holder at each irradiation for dosimetry. A motorised set-up, located in the vacuum chamber between the source and the Mylar window, was used for beam imaging and spectroscopy and was removed during irradiations. It contains a removable dipole (\SI{0.058}{T}) and a phosphor screen coupled with a CCD camera. Imaging of the beam transverse section, performed by removing the dipole magnet from the beam axis, allowed to align the electron beam to the cell sample and to measure the total accelerated charge per bunch.  The details of these diagnostics can be found in D. Gustas (2019)~\cite{gustas_high-charge_2018}.
\section{Dosimetry}
\begin{figure}[]
	\centering
	\includegraphics[width=1\columnwidth]{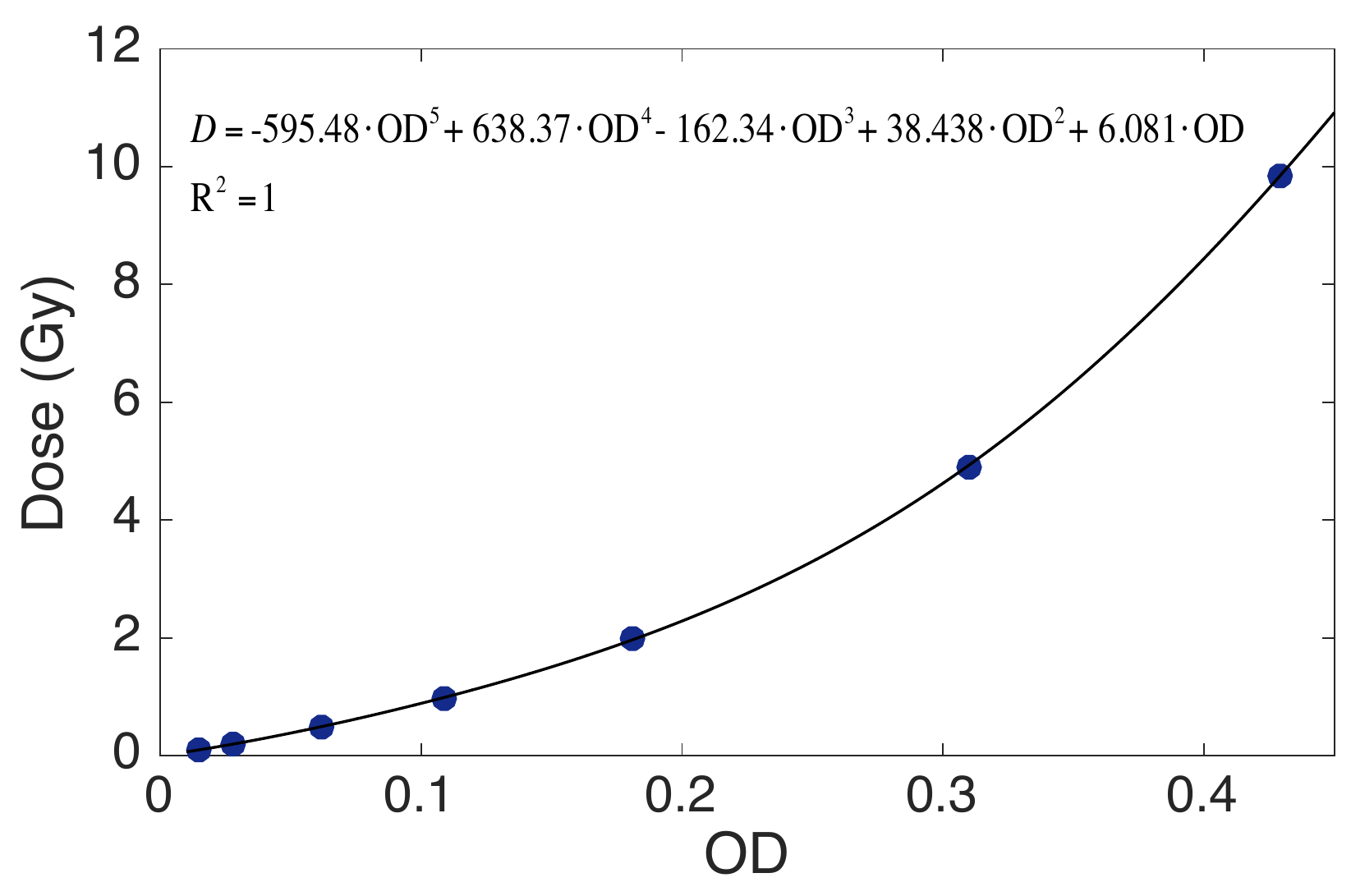}
	\caption{Calibration of the EBT3 film batch (red channel) with the \SI{4}{MeV} electron beam generated by the Synergy Elekta LINAC available at the University Hospital of Lausanne (CHUV, Switzerland). The experimental values (blue points) are fitted with a five degree polynomial (black curve), whose expression is reported in the image.}
	\label{calib ebt3 sallenoire}
\end{figure}

\begin{figure}[]
	\centering
	\includegraphics[width=1\columnwidth]{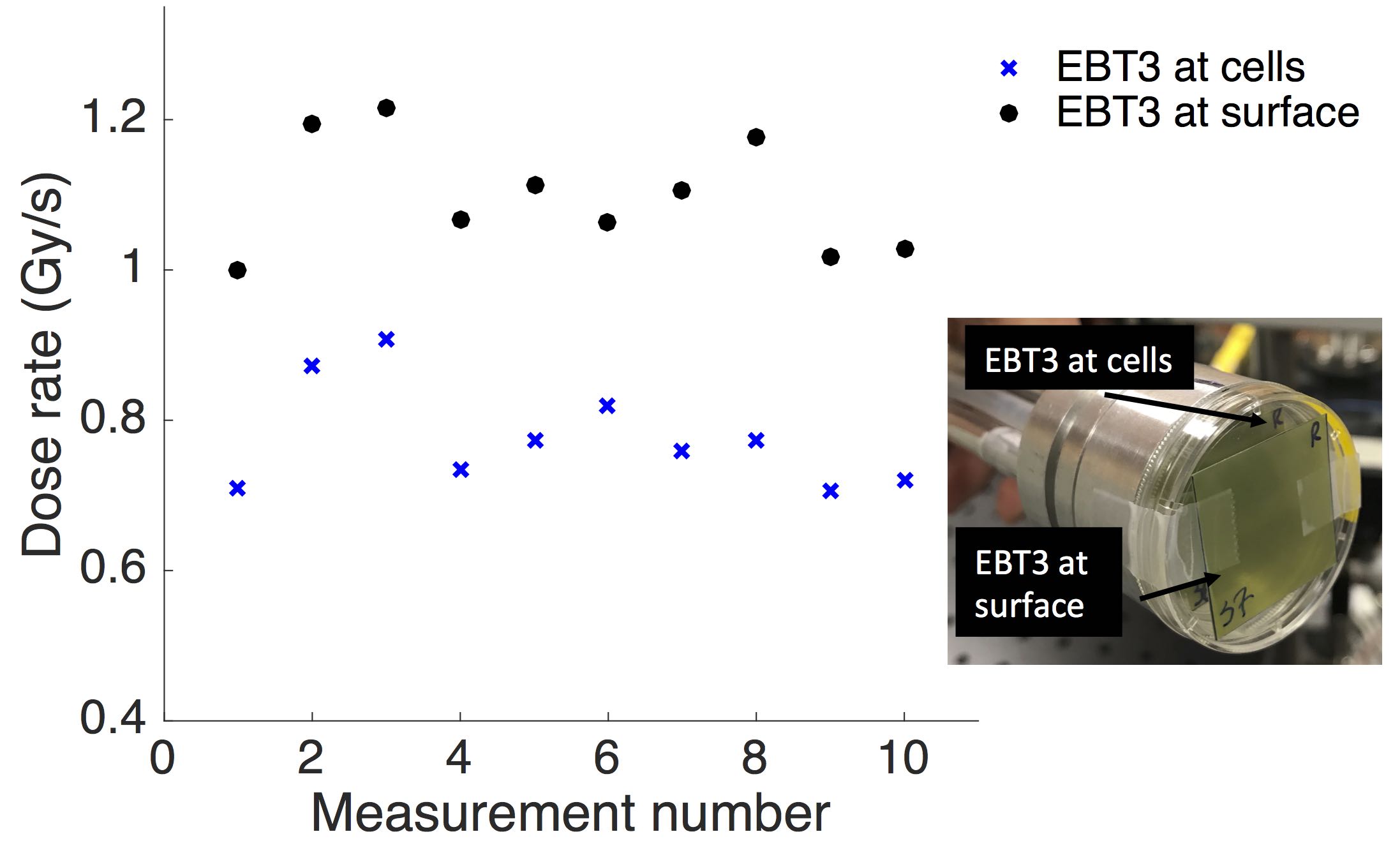}
	\caption{Dose rate measured at the surface of the plastic holder and at the cell position with the double film arrangement shown in the insert.}
	\label{double film dose}
\end{figure}

\sloppy
The dose was measured with an EBT3 radiochromic film placed on the cell plastic support at each irradiation. The films were scanned 24~hours after irradiation, as recommended by Devic et al. (2016)~\cite{devic_reference_2016}, with an EPSON V800 flatbed scanner and a resolution of 300 DPI. The dose was measured within a circular ROI of \SI{10}{mm} diameter aligned with the cell sample. The EBT3 film batch was calibrated at the Elekta Synergy linear accelerator of the \textit{University Hospital of Lausanne} (CHUV, Switzerland) with \SI{4}{MeV} electrons at a dose rate of \SI{6}{Gy/min}. The calibration curve is shown in Fig.~\ref{calib ebt3 sallenoire}, together with the five degree polynomial expression used to fit the experimental points.
\\
The film dose was corrected to obtain the dose delivered in the cells by measuring the attenuation caused by the \SI{800}{\micro\meter} cell plastic holder and the air gap. Precisely, the dose delivered to the cell is related to the dose absorbed by the film according to:
\begin{equation}
D_{cells}=D_{EBT3}\cdot R_{cells/EBT3}
\end{equation}
where \(R_{cells/EBT3}\) is the ratio between the dose absorbed by the cells and by the EBT3. This factor was measured with a double film arrangement, as shown in the insert of Fig.~\ref{double film dose}. A film was placed on the plastic holder surface, as during irradiations, and a second film was placed inside the plastic holder at the cell position. Since the ratio depends on the spectral shape of the beam, which varied slightly during the day and between days (see Fig.~\ref{spectra}), the double film measurement was carried out before each irradiation series. The results of the double film measurements, performed in five different days, are reported in Fig.~\ref{double film dose}. We measured a factor \(R_{cells/EBT3}\) of 0.71 with good reproducibility during the day and between days (5.8\% standard deviation). A mean dose rate of 1.1 Gy/s and 0.78 Gy/s was measured on the holder surface and at the cell position, respectively, with an instability of 8\% (standard deviation). As anticipated in the introduction, the mean dose rate achieved with such a kHz laser-driven electron source is higher than typical dose rates reported in literature with J-class lasers running at \SI{10}{Hz} (\(\sim\)Gy/min)~\cite{gizzi_laserplasma_2015,andreassi_radiobiological_2016,laschinsky_radiobiological_2012,oppelt_comparison_2015,beyreuther_establishment_2010,nicolai_realizing_2014}. A rough estimate of the peak dose rate in the pulse can be obtained by considering the temporal stretch of the beam energy components between \SI{0.5}{MeV} and \SI{1}{MeV} at the cell position. The calculation leads to a pulse duration of \SI{120}{ps} and a peak dose rate in the order of \SI{e7}{Gy/s}.
\subsection{Dose homogeneity and uncertainty}
\begin{figure*}[]
	\centering
	\includegraphics[width=0.8\textwidth]{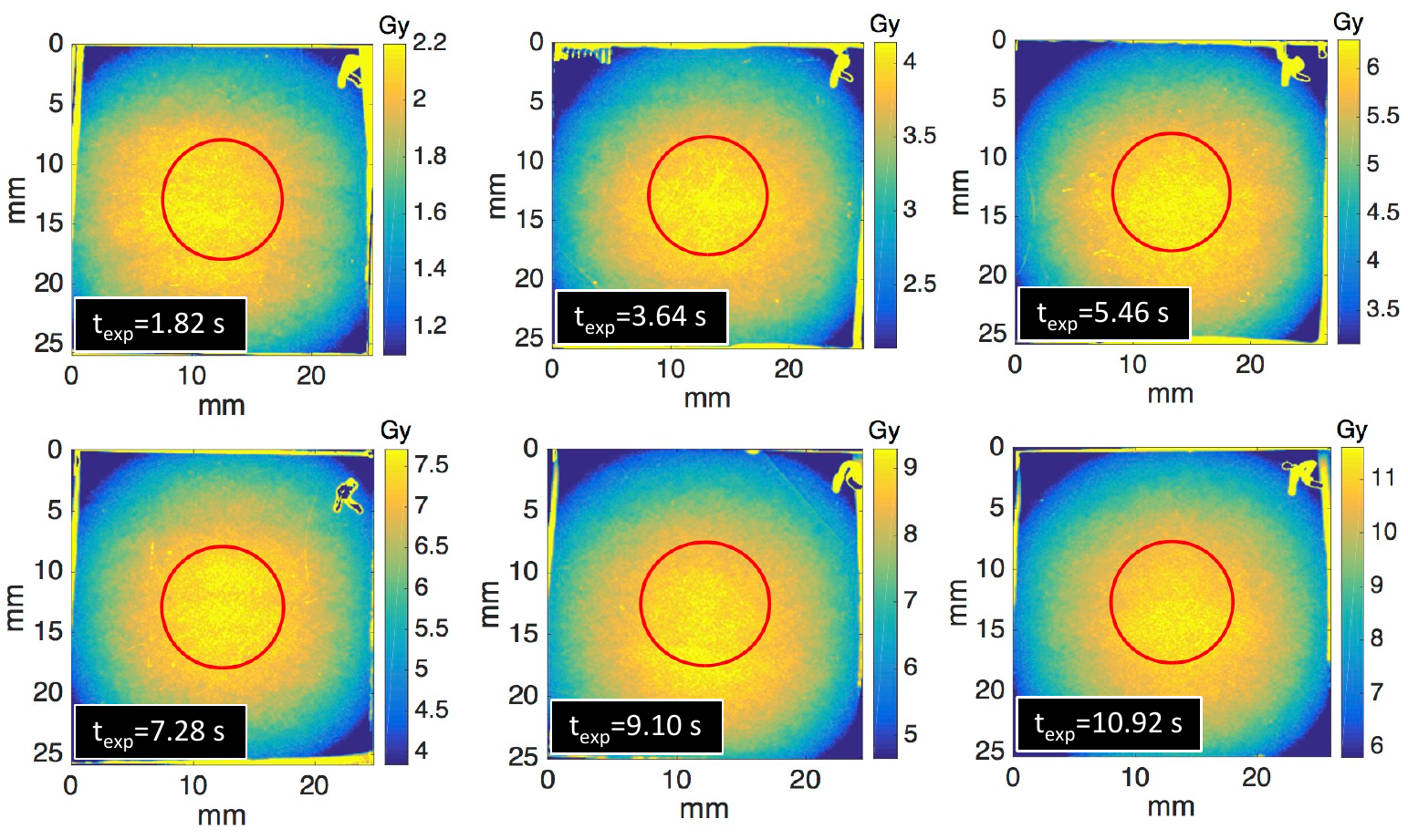}
	\caption{Dose distribution measured with the EBT3 film placed on the cell plastic support in six consecutive irradiations consisting in a dose escalation performed for the survival essay. The colormap interval is set between the maximum value and the 50\% percent level in all images. The red circles represent the ROI of \SI{1}{cm} diameter used to measure the dose as well as the surface where the cell response was studied. }
	\label{film images}
\end{figure*}

\begin{figure}[]
	\centering
	\includegraphics[width=1\columnwidth]{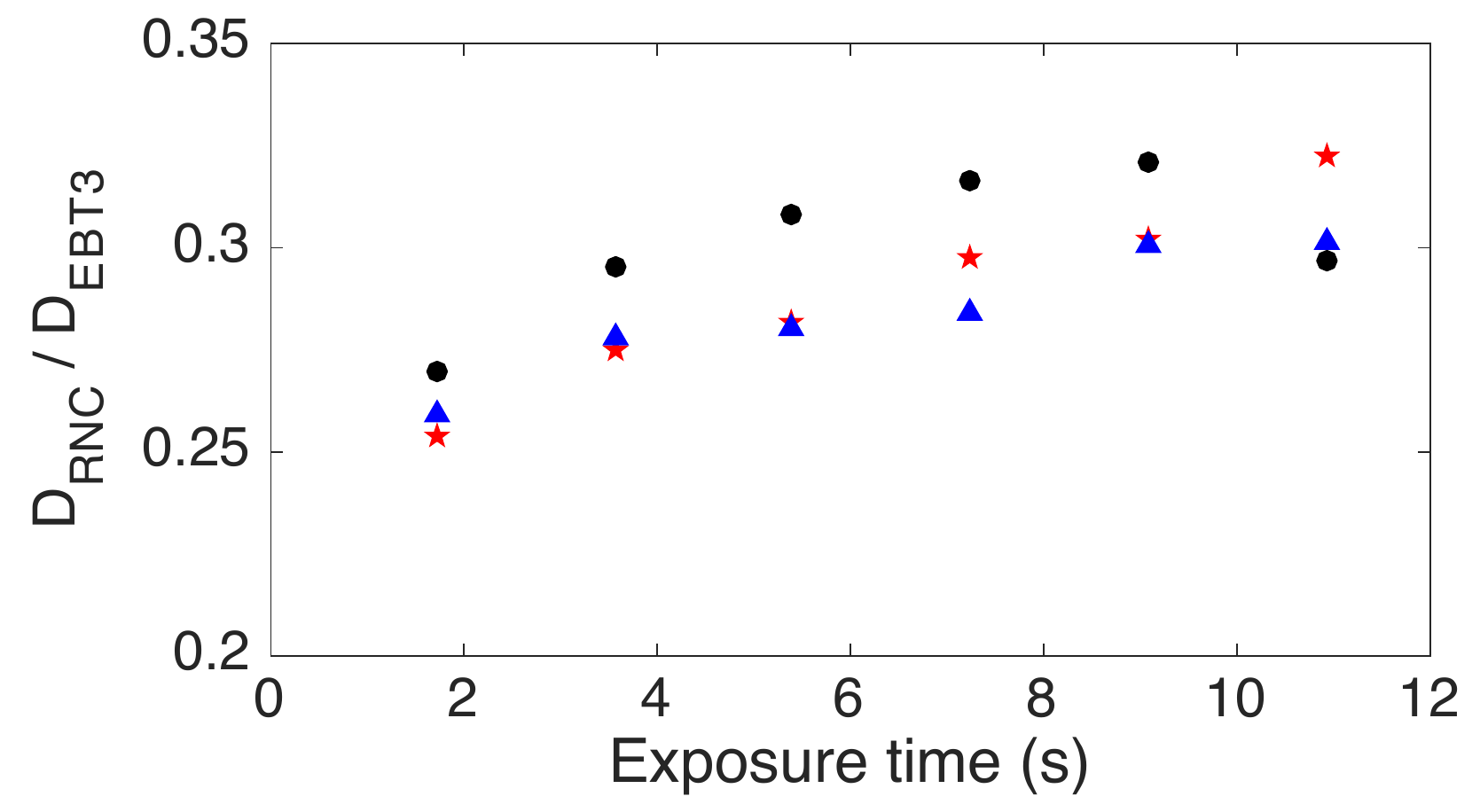}
	\caption{Ratio between the dose meaured by the EBT3 film and by the Razor Nano Chamber as a function of the exposure time, i.e. the shutter opening time.}
	\label{ratios}
\end{figure}

\sloppy
Since the position of the irradiation site inserted in the tube was fixed, the electron beam pointing needed to be carefully aligned by adjusting the incidence angle of the laser on the gas-jet before the beginning of the irradiations. Once the beam pointing was aligned with the cell holder, the beam showed a remarkably good stability in terms of dose distribution between irradiations, as shown in Fig.~\ref{film images}. 
The figure shows typical dose distributions measured with the EBT3 during a dose escalation performed for the survival assay. We were able to precisely align the electron beam with the cell sample and keep the dose inhomogeneity below \(\pm7\)\% in most of the irradiations (85\%).\\The uncertainty on the dose delivered at the cell sample was evaluated by cross correlation between the dose measured by the EBT3 film and by the RNC placed behind the sample. In particular, the instability of the ratio between the two measured doses is an indicator of the spectrum stability between irradiations and of the uncertainty on the ratio \(R_{cells/EBT3}\) used to quantify the dose delivered at cell sample. Fig.~\ref{ratios} shows the typical behaviour of the ratio between the dose absorbed by the film and by the RNC during three dose escalation series performed for the survival assay.
As it is shown, we observed an increase of the ratio with the exposure time. A similar trend was also observed for the dose rate, which turned out to slightly decrease with the exposure time. It is possible that the observed patterns are due to the shutter used to block the laser and control the irradiation time. Opening the shutter, placed before the last four optics, after it is kept closed to block the laser during the sample replacement might generate a transient thermal phenomenon in the optics and affect the laser properties at the gas-jet. This effect will be verified and solved in future experiment by placing the laser shutter after the last focusing optics. The variation of the ratio between irradiations, typically below \(\pm10\%\), results in a comparable uncertainty on the \(R_{cells/EBT3}\) factor measured with the double film arrangement and, as a consequence, on the dose delivered to the cell sample.

\section{Radiation biology experiment}
\label{sec:2}
\subsection{Methods}
\sloppy
We performed irradiation of \textit{in-vitro} cancer cells with the \textit{Salle Noire} electron source both to investigate the biological response to kHz laser-accelerated electrons and to validate the dosimetry. To this aim, we irradiated colorectal cancer cells HCT116, both the wild type (HCT116 WT) and its radioresistant counterpart (HCT116 p53\(^{-/-}\)) mutated for the tumour suppressor gene p53. The use of both the wild type and the radioresistant counterpart allowed to verify the quality of the dosimetry protocol and of the radiobiology analysis. The colorectal cancer cells HCT116, wild type (WT) or mutated for the tumour suppressor gene p53 (p53\(^{-/-}\)), were cultured and grown as monolayers in plastic tissue culture disposable flasks (TPP) in McCoy’s 5A (Modified) Medium with GlutaMAXTM (ThermoFisher Scientific), supplemented with 10\% fetal calf serum (PAA) and 1\% penicillin and streptomycin (ThermoFisher Scientific).\\All cell lineages were grown at 37\degree~C in a humidified atmosphere of 5\% CO\(_2\) in air. Based on the transverse dose homogeneity deducted from radiochromic films, a \SI{1}{cm} diameter circular region was delimited on the inside face of the culture dishes with Creamy Color long lasting lip pencil (Kiko) in which \SI{4e4}{cells} were seeded in \SI{100}{\micro\liter} of medium. The culture dish was positioned vertically behind the exit window. After exposure to the laser-driven electron beam, 1 mL of medium was added and the cell monolayers were incubated for 4~hours in standard conditions. Cells were harvested with Accutase (Merck), dispatched into 3 different wells of 12-well plate (TPP) in \SI{2.5}{\milli\liter} of medium and grown for five generations corresponding to five days (replication time below 20~h). Appropriate control samples were treated under the same conditions, including bringing the cell culture dish in a vertical position as for irradiation.\\ After this period, cells were harvested with \SI{250}{\micro\liter} of Accutase inactivated with \SI{250}{\micro\liter} of 1X PBS (ThermoFisher Scientific) supplemented with 10\% fetal calf serum. The final volume was adjusted to \SI{1}{\milli\liter} with 1X PBS and \SI{200}{\micro\liter} of each well were dispatched into a non-sterile Ubottom 96-well plate (TPP). In each well, \SI{2}{\micro\liter} of a propidium iodide solution (Sigma, \SI{100}{\micro g/\milli\liter} in 1X PBS) were added just before counting by flow cytometry technology. Cell acquisition was performed using Guava\textsuperscript{\tiny\textregistered} and the data analysis carried out with GuavaSoft (Merck Millipore).
\subsection{Survival assay results}
\sloppy
\begin{figure}[]
	\centering
	\includegraphics[width=1\columnwidth]{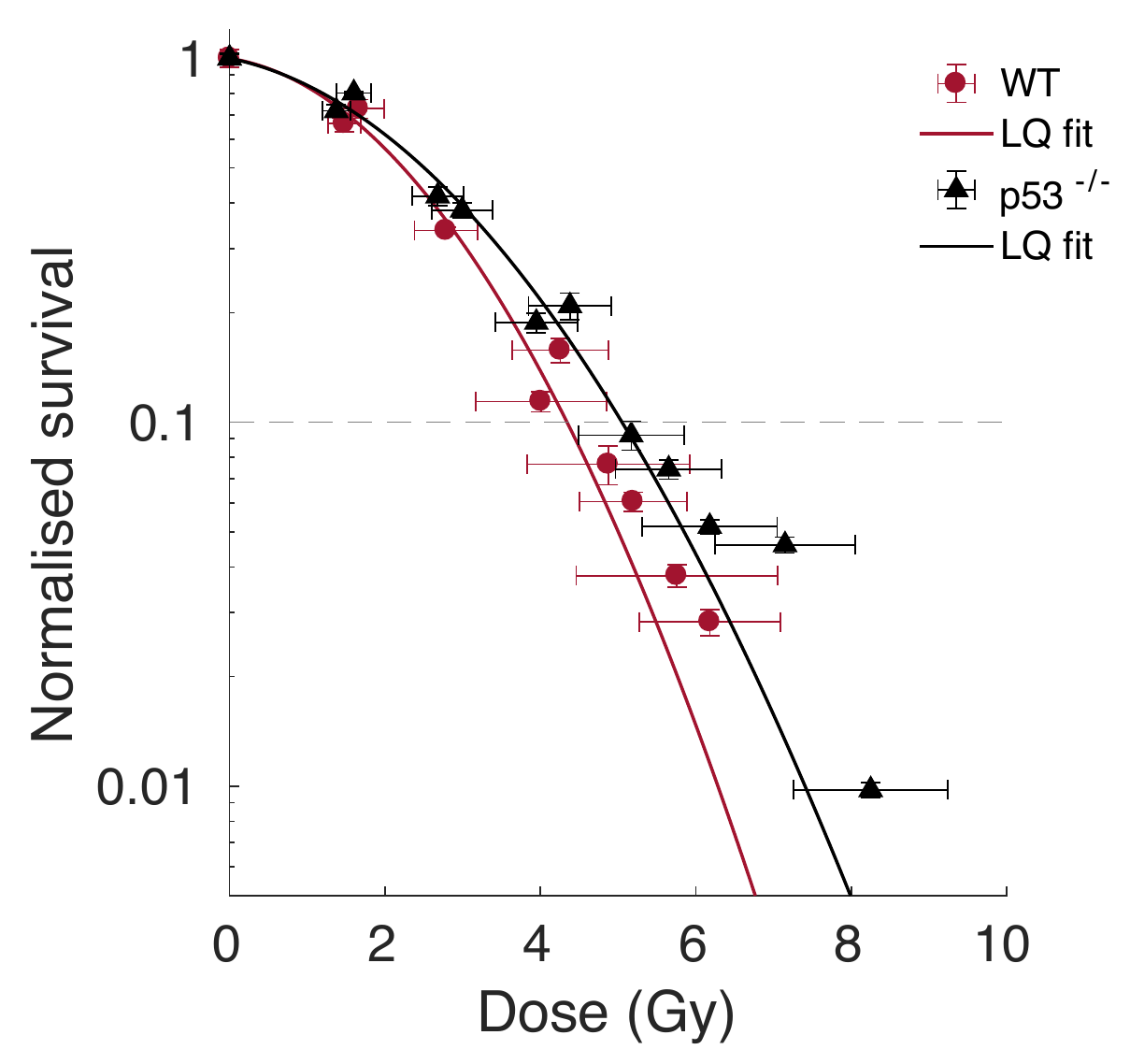}
	\caption{Survival curves for the HCT116 WT and HCT116 p53\(^{-/-}\) cell lines obtained in two independent experiments. In both graphs the points represent the mean of three radiobiological analysis carried out on the same irradiated sample and the vertical error bar corresponds to the standard deviation of the survival fraction obtained in the three analysis. The horizontal error bars represent the sum of the dose error due to the inhomogeneity within the analysed surface of the sample (circular surface of \SI{1}{cm} diameter) and of the uncertainty on the dosimetry.}
	\label{survival}
\end{figure}
The survival curves obtained for the HCT116 WT and HCT116 p53\(^{-/-}\) cell lines grown as monolayers are reported in Fig.\ref{survival}. Both the error due to the dose homogeneity and to the dosimetry uncertainty are included in the horizontal error bars. The experimental results were fitted with the Linear-Quadratic (LQ) model: \(\ln S= \alpha D-\beta D^2\), where D is the delivered dose and \(\alpha\) and \(\beta\) are adjustable parameters. The surviving fraction decreases with the delivered dose and the curves exhibit the typical shoulder shape of low LET radiations. Furthermore, the curve of the WT cell line lies well below the curve of the more radioresistant counterpart p53\(^{-/-}\), which confirms the quality of both the dosimetry and the biological analysis. 
\begin{table}[]
	\centering
	\caption{D\(_{10}\) values for \textit{in-vitro} HCT116 cell lines, grown as monolayers, obtained with laser-accelerated electrons (from the survival curves of Fig.~\ref{survival}) and with \SI{662}{keV} photons of a \(^{137}\)Cs source at a conventional dose-rate, taken from Pommarel et al. (2017)\cite{pommarel_spectral_2017}.}
	\label{D10}
	\renewcommand{\arraystretch}{1.5} 
	\begin{tabular}{c|cc}
		Cell line   & 
		LAE & Photons \(^{137}\)Cs 662 keV \\ \hline
		HCCT116 WT  & \(4.34\pm0.73\) Gy                                & \(3.8\pm0.4\) Gy           \\
		HCCT116 p53 & \(5.05\pm0.66\) Gy                                  & \(5\pm0.4\) Gy             \\ \hline
	\end{tabular}
\end{table}\\It is relevant to compare the 10\% survival dose D\(_{10}\) obtained with the LAE beam with those previously obtained with low-LET \SI{662}{keV} photons at a conventional dose rate of \SI{1.4}{Gy/min} produced by a \(^{137}\)Cs source (integrated in an IBL 637 n\(^\circ\)9418 device manufactured by CIS Bio)\cite{pommarel_spectral_2017}. 
The D\(_{10}\) values, determined from the LQ fitting curves (Kaleidagraph software) for the two cell types (WT and p53\(^{-/-}\)) are reported in Table~\ref{D10}. The D\(_{10}\) obtained with LAEs are close to the values
obtained with the \(^{137}\)Cs source, which is in line with the results reported in literature showing no remarkable difference in the RBE of LAEs in comparison with conventional sources of comparable LET~\cite{laschinsky_radiobiological_2012}. 
\section{Conclusions and outlook}
	\sisetup{
		range-phrase= -}
We presented the first dosimetric characterisation and application to radiation biology of a kHz laser-driven electron beam. The use of both radiochromic films and an on-line Razor Nano Chamber allowed evaluation of the dose distribution delivered at the sample at each irradiation and to evaluate the uncertainty on the delivered dose. We showed good day-to-day reproducibility of the irradiation conditions in terms of dose rate and beam quality. A dose rate of \SI{1}{Gy/s} was achieved, which is higher than typical dose rates reported in literature with J-class lasers running at \SI{10}{Hz}. Besides the increase of the mean dose rate, the high repetition rate ensured a remarkably good stability of the dose distribution at the sample and indicates that such LAE source can be a promising alternative to electrons produced by \SI{10}{Hz} lasers for radiation biology applications. During the dose escalation sequence, we observed an impact of the exposure time on the beam spectrum, which introduced an uncertainty on dosimetry. The observed trends are likely to be due to the use of a beam shutter in the laser chain generating transient thermal phenomena in the optics. This phenomenon, which is not attributable to inherent source instability, can be significantly reduced in future studies by placing the shutter at the end of the laser chain. In addition, an ongoing upgrade of the \textit{Salle Noire} facility will lead to an increase of the laser energy up to \SIrange[range-units = single]{4}{5}{mJ/pulse}. This would allow to generate more energetic electrons (with a comparable charge per bunch) and to reduce the sensitivity of the dosimetry protocol to spectral instabilities thanks to the higher penetration depth.\\Regarding the radiobiological outcome, the survival curves obtained with HCT116 monolayer samples indicate no significant difference between the radiobiological effectiveness of kHz laser-driven electrons and conventional sources with similar LET, in line with previous findings reported in literature. This result further validates both the dosimetry and the biological analysis. Future experiments should explore the response of healthy cells and of spheroids samples as well as the impact of different fractionation modalities on the biological response~\cite{bayart_fast_2019} to provide a more complete indication of the therapeutic potential of such high peak dose rate sources. Moreover, boosting the energies up to few MeV without a significant decrease of the charge per bunch would allow to irradiate \textit{in vivo} biological targets, such as mice, and would increase the peak dose rate in the pulse by reducing the pulse temporal stretch to few picoseconds.

\bibliographystyle{./bibliography/unsrt-5aut}
\bibliography{./bibliography/bibliography}

\end{document}